\documentclass[aps,prl,reprint,groupedaddress,showpacs]{revtex4-1}
\usepackage{amsmath, amssymb}
\usepackage{graphicx}
\usepackage{bm}

\begin{document}

\def\x{\boldsymbol{x}}
\def\E{\boldsymbol{E}}
\def\rot{\boldsymbol{\nabla} \times}
\def\rotp{\boldsymbol{\nabla}' \times}
\def\om{\omega}
\def\ep{\varepsilon}


\title{Analysis in temporal regime of dispersive invisible structures designed from transformation optics}



\author{B. Gralak}
\email{boris.gralak@fresnel.fr}
\affiliation{CNRS, Aix-Marseille Universit{\'e}, Centrale Marseille, Institut Fresnel, 13397 Marseille, France}
\author{G. Arismendi}
\affiliation{CNRS, Aix-Marseille Universit{\'e}, Centrale Marseille, Institut Fresnel, 13397 Marseille, France}
\author{B. Avril}
\affiliation{CNRS, Aix-Marseille Universit{\'e}, Centrale Marseille, Institut Fresnel, 13397 Marseille, France}
\author{A. Diatta}
\affiliation{CNRS, Aix-Marseille Universit{\'e}, Centrale Marseille, Institut Fresnel, 13397 Marseille, France}
\author{S. Guenneau}
\affiliation{CNRS, Aix-Marseille Universit{\'e}, Centrale Marseille, Institut Fresnel, 13397 Marseille, France}

\date{\today}

\begin{abstract}
A simple invisible structure made of two anisotropic homogeneous layers is analyzed 
theoretically in temporal regime. The frequency dispersion is introduced and 
analytic expression of the transient part of the field is derived for large 
times when the structure is illuminated by a causal excitation. This 
expression shows that the limiting amplitude principle applies with transient 
fields decaying as the power $-3/4$ of the time. The quality of the cloak is 
then reduced at short times and remains preserved at large times. 
The one-dimensional theoretical analysis is supplemented with
full-wave numerical simulations in two-dimensional situations which 
confirm the effect of dispersion.
\end{abstract}


\maketitle


In 2006, Pendry et al. 
\cite{Pendry06} and Leonhardt
\cite{Leonhardt06} designed
an invisibility cloak for electromagnetic radiation by blowing up a 
hole in optical space and hiding an object inside it. 
These proposals have been validated by microwave experiments \cite{Smith06}.
However, these metamaterials are subject to an 
inherent frequency dispersion which may affect the quality of the optical 
function designed in time harmonic regime. Hence, there is a renewed interest 
in the propagation in dispersive media, originally investigated by Brillouin 
\cite{Bri1960}. 
The effect of dispersion has been addressed in the cases of the flat lens 
\cite{Collin10,GT10,GM12,PRL-Pen11,PRL-Gref12} and cylindrical invisibility 
cloaks \cite{disp2008, disp2010, Rajput:15}.


In the present letter, a regularized version of Pendry's transform 
\cite{Kohn08}, is implemented 
for the design of the simplest possible system of invisible layers.
With this transform, infinities are avoided in the material parameters of the 
cloak which consists of two homogeneous anisotropic slabs. Frequency 
dispersion is introduced, which is a required model for metamaterials whenever 
the permittivity (or permeability) is lower than that of vacuum (i.e. when the
phase velocity is greater than $c$ or negative). The effect of dispersion is 
analyzed with electromagnetic sources with sinusoidal time dependence 
that are switched on at an initial time. Such an illumination has been originally 
used by Brillouin \cite{Bri1960} in homogeneous dispersive media, and more 
recently in the case of the negative index flat lens \cite{Collin10,GT10,GM12}. 

The originality of our approach is to consider a simple invibility 
system made of two layers allowing analytic calculations. Indeed, the invisible 
nature of the system leads to a simple expression 
of the transmitted field, since there is no reflexion at the interfaces. 
Also, the absence of branch cut in the integral expression of the 
time dependent field in multilayered structures is exploited. 
The method is presented in detail 
and the derivation of the 
transient regime shows that the electromagnetic field includes contributions 
generated by the singular values of the permittivity and permeability (zeros 
and infinities). An explicit expression of the transient fields is obtained 
for long times, which is similar to the one obtained by Brillouin 
\cite{Bri1960} for wavefronts (forerunners). Next, the limiting amplitude 
principle is considered to show 
that cloaking can be addressed in temporal regime after the transient regime.  
These results are supplemented with numerical simulations in the case of a 
two-dimensional cylindrical layered cloak, where the presence of additional modes 
is confirmed in the transient regime.

\begin{figure}[b]
\centering
\fbox{\includegraphics[width=0.96\linewidth]{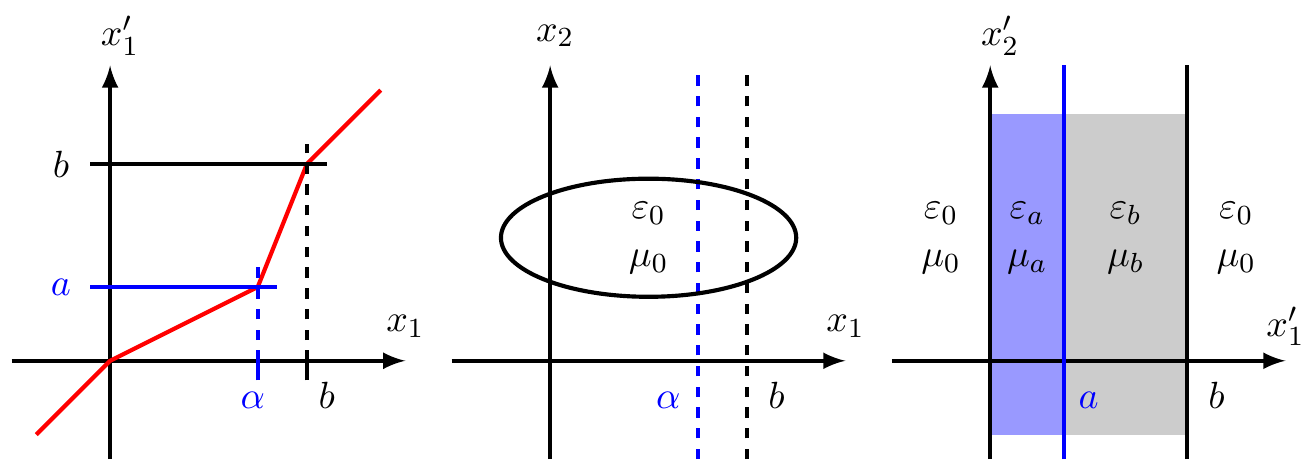}}
\caption{Coordinate transform for invisible layers.  
Left: change of coordinate $x_1 \rightarrow x_1'$.
Center: free space before coordinate transform.
Right: invisible set of homogeneous anisotropic layers 
after coordinate transform.}
\label{fig1}
\end{figure}

We start with the definition of a system of invisible layers. Let 
$\x = (x_1,x_2,x_3)$ be a Cartesian coordinate system 
in the space $\mathbb{R}^3$. At the oscillating
frequency $\om$, the electric field amplitude $\E(\x)$ is
governed in free space by the Helmholtz equation
\begin{equation} 
- \rot \rot \E(\x) + \om^2 \mu_0 \ep_0 \, \E(\x) = \boldsymbol{0} \, ,
\label{Helm}
\end{equation}
where $\ep_0$ and $\mu_0$ are the vacuum permittivity and permeability. 
The invisible layered structure is then deduced using the 
coordinate transform $\x \rightarrow \x'$ (see Fig. \ref{fig1}):
\begin{equation}
\begin{array}{ll}
x_1' = \dfrac{a}{\alpha} \, x_1 \quad \quad & 0 \leq x_1 \leq \alpha \, , \\[0mm] 
x_1' = a + \dfrac{b-a}{b-\alpha} \, (x_1 -\alpha) \quad \quad & \alpha \leq x_1 \leq b \, , \\[0mm] 
x_1' = x_1 \quad \quad & x_1 \leq 0 \, , \quad b \leq x_1 \, , 
\end{array}
\end{equation}
where $0 < a < \alpha < b$, $x_2' = x_2$ and $x_3' = x_3$ being 
invariant. 
The effect of this geometric transform is to map the layer 
$0 \leq x_1 \leq \alpha$ onto the layer $0 \leq x_1' \leq a$ 
(denominated as layer A), and the layer $\alpha \leq x_1 \leq b$ 
onto $a \leq x_1' \leq b$ (denominated as layer B). Note that such a geometric 
transform, adapted from \cite{Kohn10}, regularizes the original transform 
for an invisibility cloak proposed in \cite{Pendry06}. The corresponding 
transform is applied to the Helmholtz equation (\ref{Helm}):
\begin{equation} 
- \rotp \dfrac{1}{\mu(x_1')} \rotp \E'(\x') + \ep(x_1') \, \om^2 \mu_0 \ep_0 \, \E'(\x') = \boldsymbol{0} \, ,
\label{Helm2}
\end{equation}
where the relative permittivity and permeability
are both equal to the tensor $\nu \equiv \ep = \mu$ (as in \cite{PRL-Pen11}) 
taking constant values in each layer:
\begin{equation}
\left\{
\begin{array}{ll}
\ep(x_1') = \mu(x_1') = \nu(x_1') = \nu_a \quad & \text{ if } \quad 0 \leq x_1' \leq a \, , \\[0mm] 
\ep(x_1') = \mu(x_1') = \nu(x_1') = \nu_b \quad & \text{ if } \quad a \leq x_1' \leq b \, , \\[0mm] 
\ep(x_1') = \mu(x_1') = \nu(x_1') = 1 \quad & \text{ if } \quad  x_1' \leq 0 \, , \quad\!\! b \leq x_1' \, . 
\end{array}
\right.
\end{equation}
The constant values in layers A and B are given by
\begin{equation}
\nu_{a,b} = \left[ \begin{array} {ccc}
\!\nu_{a,b}^{\perp}\! & 0 & 0 \\[0mm] 0 & \!\nu_{a,b}^{\parallel}\! & 0 \\[0mm] 0 & 0 & 
\!\nu_{a,b}^{\parallel}\! \end{array} \right] , \quad 
\label{slab-a-b}
\end{equation}
where the components parallel and perpendicular to the plane interfaces, respectively 
denoted by the superscripts $\parallel$ and $\perp$, are
\begin{equation}
\nu_{a}^{\perp} = 1/ \nu_{a}^{\parallel} = a / \alpha \, , \quad \quad 
\nu_{b}^{\perp} = 1/ \nu_{b}^{\parallel} = (b-a) / (b-\alpha) \, .
\end{equation}
The transformed Helmholtz equation (\ref{Helm2}) can be reduced to a set of two independent scalar 
equations using the symmetries of the geometry, namely the invariances under the translations 
and rotations in the plane $(x_2',x_3')$. After a Fourier decomposition from $(x_2',x_3')$ 
to $(k_2',k_3')$, equation (\ref{Helm2}) becomes
\begin{equation}
\dfrac{\partial}{\partial x} \, \dfrac{1}{\nu^{\parallel}(x)} \, 
\dfrac{\partial U}{\partial x} (x) - \dfrac{k^2}{\nu^{\perp}(x)} U(x) + 
\dfrac{\om^2}{c^2} \nu^{\parallel}(x) U(x) = 0 \, ,
\label{Hscalaire}
\end{equation}
for $U(x)$, the (Fourier transformed) electric field component along direction 
$(-k_3,k_2)$. Here, $x$ denotes $x_1'$, $k^2$ equals $k_2^2 + k_3^2$ (with 
$k_2 = k_2'$ and $k_3 = k_3'$), $c = 1/\sqrt{\ep_0 \mu_0}$ is the light velocity in 
vacuum, and functions $\nu^{\parallel}(x)$ and $\nu^{\perp}(x)$ 
are the components of $\nu(x)$ respectively parallel and perpendicular to the 
plane interfaces. Notice that, since $\ep = \mu$, the second scalar equation derived 
from the Helmholtz equation is fully identical to (\ref{Hscalaire}), except that 
$U(x)$ should be the (Fourier transformed) magnetic field component along direction 
$(-k_3',k_2')$ [or $(-k_3,k_2) $].

In this letter, the system is analyzed using a transfer matrix formalism \cite{Liu13}. 
Equation (\ref{Hscalaire}) is formulated as 
\begin{equation}
\dfrac{\partial}{\partial x} \, F(x) = - i M(x) F (x) \, ,
\label{dF}
\end{equation}
where 
\begin{equation}
F = \left[ \begin{array}{c} U \\ \dfrac{i}{\nu^{\parallel} } \, \dfrac{\partial U}{\partial x} 
\end{array} \right] \, , \quad 
M  = \left[ \begin{array}{lr} 0 & \nu^\parallel  \\ \dfrac{\om^2}{c^2} \nu^{\parallel} 
- \dfrac{k^2}{\nu^{\perp} } & 0 
\end{array} \right] \, .
\label{FM}
\end{equation}
The transfer matrices $T_{a}$ and $T_{b}$, associated with layers A and B, defined by 
$F(a) = T_a F(0)$ and $F(b) = T_b F(a)$, are given by
\begin{equation}
T_a = \exp[-i M_0 \alpha] \, , \quad T_b = \exp[- i M_0 (b - \alpha)] \, ,
\label{F}
\end{equation}
the matrix $M_0$ being the value taken by the matrix $M(x)$ in vacuum, i.e. when 
$\nu^\parallel(x) = \nu^\perp(x) = 1$. This implies that the transfer matrix 
$T_b T_a = \exp[- i M_0 b]$, associated with layers A and B, is exactly the 
same as the one of a vacuum layer of thickness $b$. Hence the system 
of layers A and B is invisible to any incident field. 

Nevertheless, as pointed out by V. Veselago when he introduced negative index 
materials \cite{Ves68}, causality principle and passivity require that permittivity 
and permeability be frequency dispersive when they take relative value below unity 
\cite{Landau,Jackson}. According to this requirement, frequency dispersion is 
introduced in the components of $\nu_a$ and $\nu_b$ with value below unity, assuming
the simple Drude-Lorentz model \cite{Jackson}:
\begin{equation}
\begin{array}{ll}
\nu_a^\perp(\omega) = 1 - \dfrac{\Omega_a^2}{\om^2 - \om_a^2} \, , \quad 
&\Omega_a^2 = \dfrac{\alpha- a}{\alpha} \, (\om_0^2 - \om_a^2) \, , \\[2mm]
\nu_b^\parallel(\omega) = 1 - \dfrac{\Omega_b^2}{\om^2 - \om_b^2} \, , \quad 
&\Omega_b^2 = \dfrac{\alpha - a}{b-a} \, (\om_0^2 - \om_b^2) \, .
\end{array}
\label{disp}
\end{equation}
Under this assumption, the functions $\nu_a^\perp(\omega)$ and $\nu_b^\parallel(\omega)$ 
take the appropriate values for the invisibility at $\om = \om_0$. 
Notice that the resonance frequencies $\om_a$ and $\om_b$ must be smaller than 
the operating frequency $\om_0$ in order to ensure that the oscillator strengths $\Omega_a^2$ 
and $\Omega_b^2$ are positive. For frequencies different from $\om_0$, the system has no reason 
to be invisible. 


The effect of dispersion is analyzed using illumination with sinusoidal time-dependence 
oscillating at $\om_0$ and switched on at an initial time. Such a ``causal'' incident 
field, originally used by L. Brillouin \cite{Bri1960} and more 
recently in \cite{Collin10,GT10,GM12}, is assumed to be in normal 
incidence for simplicity. Hence the following current source is considered: 
\begin{equation}
S(x,t) = S_0 \, \delta(x - x_0) \theta(t) \, \sin[\om_0 t] \, , 
\label{source}
\end{equation}
where $\delta$ is the Dirac ``function'', $\theta(t)$ the step function 
(equal to 0 if $t<0$ and 1 otherwise), and $S_0$ the constant component 
of the source parallel to the field component $U(x)$. 
\begin{figure}[h]
\centering
\fbox{\includegraphics[width=0.96\linewidth]{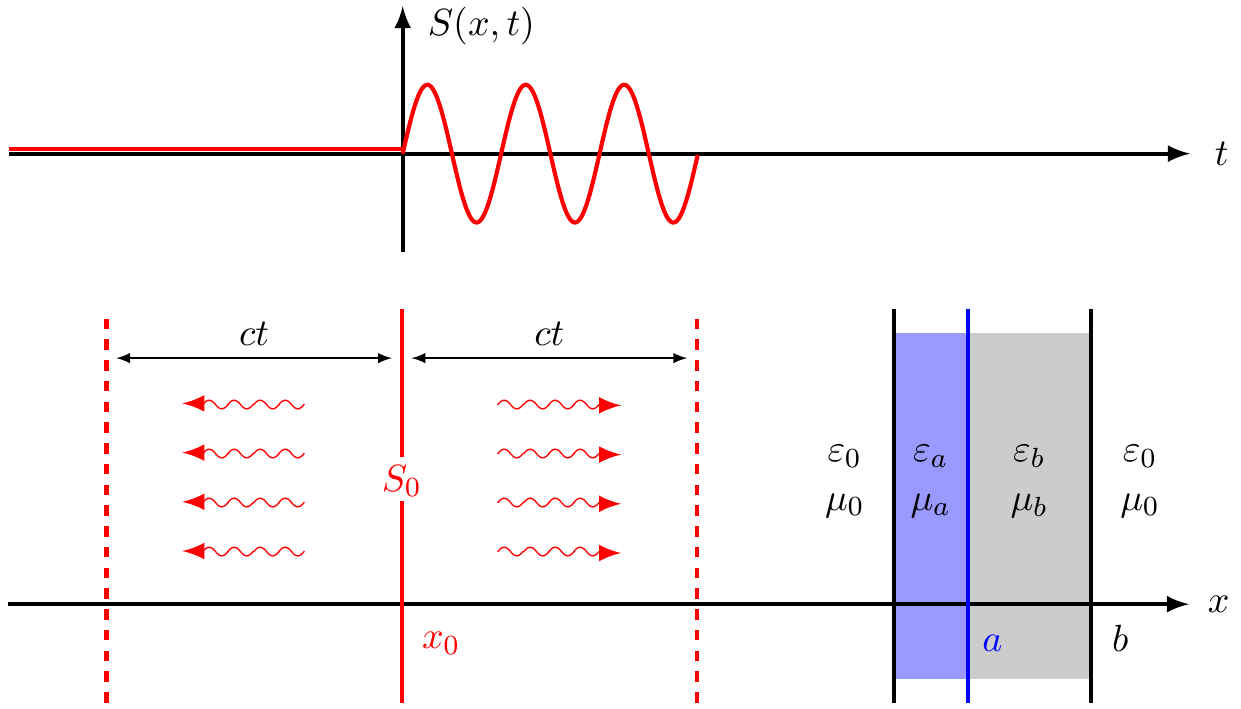}}
\caption{Excitation of the system.  
Top: Causal current source with sinusoidal time dependence.
Bottom: Field radiated by the causal source and illuminating the 
invisible layers.}
\label{fig2}
\end{figure}
In the domain of complex 
frequencies $z = \om + i \eta$, 
the electric field radiated in vacuum by this source is 
\begin{equation}
U_0(x,z) = \dfrac{ S_0 \mu_0 c }{2} \, \dfrac{\om_0}{z^2 - \om_0^2} \, 
\exp\big[i z |x - x_0| / c \big] \, .
\label{Uinc}
\end{equation}
The positive imaginary part $\eta$ has been added to the frequency $\om$
to ensure a correct definition of the Fourier transform with respect to time 
of the source (\ref{source}). The time dependent incident field radiated in 
vacuum is, with 
$z=\om+i\eta$,
\begin{equation}
\begin{array}{lcl}
E_0(x,t) & = & \dfrac{1}{2\pi} 
\displaystyle\int_{\mathbb{R}} d\om \, \exp[- iz t] \, 
U_0(x,z) \, \\[3mm]
& = & - \dfrac{ S_0 \mu_0 c }{2} \, \theta(t - |x - x_0| / c) \, \\[3mm]
& & \times \sin[\om_0 (t - |x - x_0| / c)] \, .
\end{array}
\label{incField}
\end{equation}
The next steps are to compute the time dependent field transmitted through 
the system, and to analyze the behavior of the field when the time $t$ 
tends to infinity. According to the limiting amplitude principle, the solution 
should have an asymptotic behavior corresponding to the time 
harmonic frame oscillating at the frequency $\om_0$. 
Let $T(\om)$ be the transmission coefficient of the system made of layers 
A and B. Then, the time dependent electric field is, for $x>b$, 
\begin{equation}
E_T(x,t) = \dfrac{1}{2\pi} 
\displaystyle\int_{\mathbb{R}} d\om \, \exp[- iz (t - \{x - b\} / c) ] \, 
U_0(0,z) \, T(z) \, .
\label{TField}
\end{equation}
At this stage, it is stressed that, for a fixed incident angle, 
the transmission coefficient $T(z)$ does not contain any square root of the 
permittivities and permeabilities of the layered system and of the complex 
frequency $z$. This remarkable property, which remains true for any multilayered 
structure, underpins the present technique
since it removes all branch cuts in the evaluation of the integral of the transmitted 
field. This is an advantage in comparison with the method
used by Brillouin for the analysis of wave propagation in dispersive media 
\cite{Bri1960}. The expression of the transmitted field 
is thus given by the sum of the contributions from all the poles in the function 
$f(z)$ under the integral in (\ref{TField}).

The poles of the factor $U_0(0,z)$ at $z = \pm \om_0$ [see Eq. (\ref{Uinc})] 
provide the contribution at the operating frequency $\om_0$,
\begin{equation}
\begin{array}{ll}
E_T^{(0)}(x,t) = & - \dfrac{ S_0 \mu_0 c }{2} \, 
\theta(t - \{x-x_0 + \alpha -a \} / c) \\[2mm]
& \times \sin[\om_0 (t - \{x-x_0\} / c)] \, ,
\end{array}
\label{TField0}
\end{equation}
corresponding to the time harmonic solution for which the system is invisible. 
This contribution vanishes for times such that $ct$ is smaller than 
$x - x_0 + \alpha - a = x + |x_0| + \alpha - a > x + |x_0|$, 
instead of $x - x_0 = x + |x_0|$. This is not suprising since the dispersion has 
not been taken into account in both parallel permittivity and permeability 
$\ep_a^\parallel = \mu_a^\parallel = \nu_a^\parallel > 1$ of layer A: 
hence the corresponding delay $(\alpha-a)/c$ is retrieved in the above 
expression. 

The poles of the transmission coefficient are determined from 
the expression
\begin{equation}
T(z) = \exp[i z \{ \alpha +  (b-a) \nu_b^\parallel(z) \} / c] \, .
\label{Tlayers}
\end{equation}
%
Next, replacing $\nu_b^\parallel(z)$ by the dispersive model (\ref{disp}) yields
\begin{equation}
T(z) = \exp[i z (\alpha + b-a) / c ] \, 
\exp \left[ - i \, \dfrac{z (b-a)}{c} \dfrac{\Omega_b^2}{z^2 - \om_b^2} \right] \, .
\label{Tdisp}
\end{equation}
Thus the transmission coefficient has two isolated singularities at $z = \pm \om_b$. 
It is shown in the supplemental material that the residues associated with these 
singularities exist, and can be estimated for large values of the relative time
\begin{equation}
\tau = t - \dfrac{x-x_0 + \alpha -a }{c} \gg \beta = 
\dfrac{(b-a) \Omega_b^2}{2\om_b^2 c} \, .
\label{tau}
\end{equation}
The resulting contribution $E_T^{(b)}$ in the transmitted field is 
\begin{equation}
\begin{array}{ll}
E_T^{(b)}(x,t) \underset{\tau/\beta\rightarrow\infty}{\approx} 
 & - \, 2 S_0 \mu_0 \pi c \, 
\dfrac{\om_0 \om_b }{\om_b^2 - \om_0^2} \, \theta(\tau) \, \dfrac{1}{\sqrt{\tau/\beta}} \, 
\\[1mm]
& \times J_1(2 \om_b \beta\sqrt{\tau/\beta}) 
\cos[\om_b(\tau + \beta / 2)\big] \, ,\\[2mm]
\end{array}
\label{ETbJ1}
\end{equation}
where $J_1$ is the Bessel function (see the supplemental material).
It is stressed that a similar behavior, given by the Bessel function 
$J_1$ with argument proportional to $\sqrt{\tau}$, has been highlighted 
by Brillouin \cite{Bri1960} but for short relative time $\tau$ 
(forerunners). In both 
cases, $J_1$ is a consequence of the 
dispersion given by the Drude-Lorentz model (\ref{disp}), but for different 
frequency ranges: near the resonance frequencies $\pm \om_b$ in the present 
case, and for the high frequencies in the case considered by Brillouin 
(forerunners). Forerunners at $\tau \rightarrow 0$ can be also 
characterized here. 

The asymptotic form $J_1(u) \approx \sqrt{2/(\pi u)} \cos[u - 3 \pi/4]$ 
provides an explicit expression for long time 
$\tau \gg \beta$. The contribution in the transmitted field becomes 
\begin{equation}
\begin{array}{ll}
E_T^{(b)}(x,t) \underset{\tau/\beta\rightarrow\infty}{\approx} 
& - \, 2 S_0 \mu_0 c \, 
\dfrac{\om_0 \om_b }{\om_b^2 - \om_0^2} \, \dfrac{\sqrt{\pi}}{\sqrt{\om_b \beta}}
\, \theta(\tau) \\[4mm]
& \times \, (\tau/\beta)^{-3/4}\, \cos\big[2 \om_b \beta \sqrt{ \tau/\beta} - 3 \pi / 4] \\[3mm]
& \times \cos[\om_b\beta (\tau/\beta + 1/2)\big] .
\end{array}
\label{ETbfinal}
\end{equation}
This expression shows that this second contribution has 
a first factor oscillating at the frequency $\om_b$ and a second 
factor with more complex oscillating behavior with argument 
$\Omega_b \sqrt{ 2 (b-a) \tau /c}$. The amplitude 
of this contribution decreases like $(\om_b \tau)^{-3/4}$, 
and thus the total transmitted electric field 
\begin{equation}
E_T(x,t) \underset{\tau/\beta\rightarrow\infty}{\approx} 
- \dfrac{S_0 \mu_0 c}{2} \, \theta(\tau) \, \sin[\om_0 (\tau + \{ \alpha - a \}/c )] 
\label{TFieldinfty}
\end{equation}
tends to the field radiated in vacuum (\ref{incField}) for long enough time 
$\tau$, and cloaking is addressed. Hence the limiting amplitude 
principle applies here, unlike for the perfect lens \cite{Collin10,GM12}. 

The situation where small absorption is included can be considered: the 
resonance frequencies $\pm \om_b$ are replaced by $\pm \om_b - i \gamma$ 
with $\gamma >0$ in (\ref{disp}) while $\Omega_b$ remains positive. 
The main change in the second contribution (\ref{ETbfinal}) is the 
presence of the additional factor $\exp[- \gamma \tau]$, which makes the 
permanent regime (purely oscillating at the operating frequency $\om_0$) 
easier to handle. Notice that the argument of the 
Bessel function, $2 \om_b \beta\sqrt{\tau/\beta} = \Omega_b \sqrt{ 2 (b-a) \tau /c}$, 
is independent of $\om_b$ 
and thus absorption has no influence on
the behavior governed by this function. Finally, it is stressed that the 
introduction of small absorption affects the transmission coefficient at 
the operating frequency $\om_0$ by an attenuation of 
$\exp[- \gamma (b-a)/c]$, which results in a signature of the invisible
structure. 

In oblique incidence, expressions are more complicated 
since reflections occur at the different interfaces. However, 
the term $- k^2/\nu^\perp_a$ in (\ref{FM}) leads to a singularity 
at the frequency $\om_p$ for which $\nu_a^\perp$ vanishes:
\begin{equation}
\nu^\perp_a(\om_p) = 0 \, , \quad \om_p = \pm \sqrt{\om_a^2 + \Omega_a^2} \, .
\end{equation}
This singularity generates an additional contribution at the 
frequency $\om_p$, as well as the singularity at $\om_b$. 
It is found that both singularities $\nu \rightarrow 0$ and 
$\nu \rightarrow \infty$ lead to additional contributions of the field 
in temporal regime. This result confirms the well-known difficulties 
associated with cloak's singularities \cite{Kohn10}.


The analytical results are numerically tested in the 
case of a cylindrical cloak designed using homogenization techniques 
\cite{Huang:07,Greenleaf:08}. This cloak is a concentric multilayered 
structure of inner radius $R_1$ and outer radius $R_2=2 R_1$, consisting of $20$ 
homogeneous layers of equal thickness $R_1/20$ and made of non dispersive dielectrics 
(see table \ref{tab1} for the values of relative permittivities, the relative 
permeability being unity).
\begin{table}[h]
\begin{center}
\begin{tabular}{|c|c|c|c|c|c|c|c|c|c|c|}
\hline
layer & 1 & 2 & 3 & 4 & 5 & 6 & 7 & 8 & 9 & 10   \\
\hline
$\varepsilon/\varepsilon_0$ & $0.0012$ & 8.0 & $0.02$ & 8.0 & $0.07$ & 8.0 
& $0.12$  & 8.0 & $0.18$ & 8.0 \\
\hline
\hline
layer & 11 & 12 & 13 & 14  & 15 & 16 & 17 & 18 & 19 & 20 \\
\hline
$\varepsilon/\varepsilon_0$ & $0.24$ & 8.0 & $0.3$ & 8.0 & $0.38$ & 8.0 & $0.44$ & 8.0 & $0.5$ & 8.0 \\
\hline
\end{tabular}
\caption{Relative permittivity values of the layered cloak from inside
(layer 1) to outside (layer 20). 
}
\label{tab1}
\end{center}
\end{table} 

\begin{figure}[h]
\centering
\fbox{\hspace*{2mm}\includegraphics[width=0.92\linewidth]{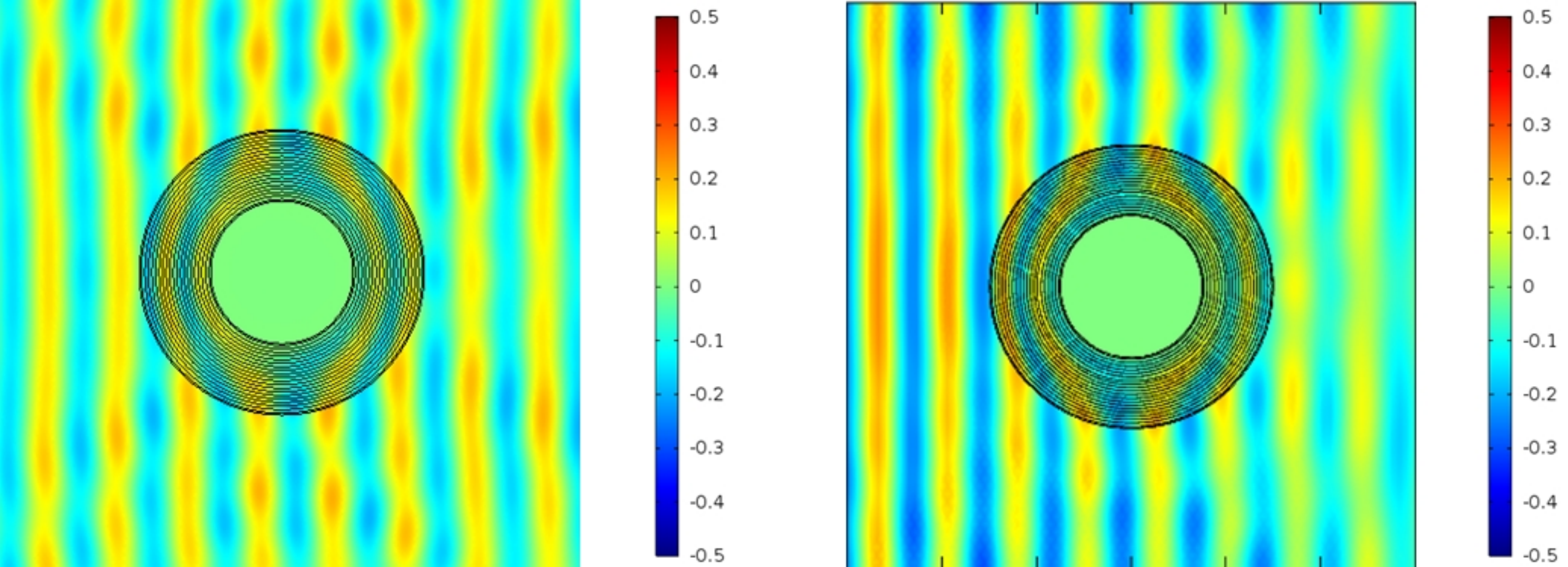}\hspace*{2mm}}
\caption{Magnetic field in the presence of the cylindrical 
cloak when illuminated by a time harmonic plane wave (left) and by 
the causal incident field given by Eq. (\ref{source}) and Fig. \ref{fig2} (right).
}
\label{fig3}
\end{figure}

The left panel of Fig. \ref{fig3} shows that the cylindrical cloak works 
almost perfectly in time harmonic regime oscillating at the frequency 
$\om_0=2\pi c/\lambda_0$, where $\lambda_0=R_2/2$. Note 
that a purely dielectric structure is used for this 2D cloak, and thus 
interfaces between different concentric layers are subject to reflections 
producing effective dispersion. Hence, it is expected to observe an effect 
of dispersion even if all the dielectric layers are non dispersive \cite{Liu13}.
The right panel of Fig. \ref{fig3} shows the longitudinal magnetic
field amplitude when the
cloak is illuminated by the causal incident field given by Eq. (\ref{source}) 
and Fig. \ref{fig2}.  

\begin{figure}[h]
\centering
\fbox{\hspace*{2mm}\includegraphics[width=0.92\linewidth]{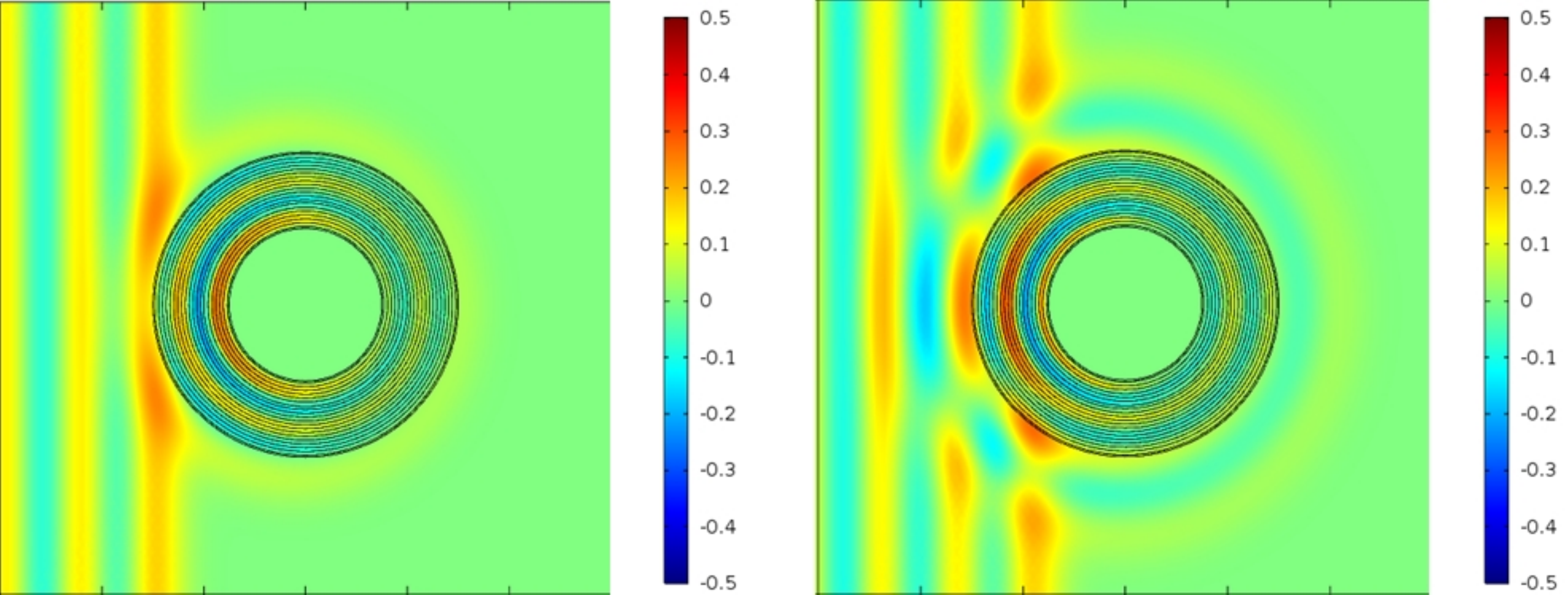}\hspace*{2mm}}
\caption{Magnetic field in the presence of the cylindrical 
cloak when illuminated by the causal incident field at two time steps
in the transient regime. Cylindrical modes inside the
cloak generate a supraluminal concentric wave.
}
\label{fig4}
\end{figure}

The cloaking effect appears to be of similar quality in both
panels of Fig. \ref{fig3}. 
We now analyze the magnetic field at short times.
In Fig. \ref{fig4}, cylindrical modes are excited in the multilayers when the 
incident front 
wave reaches the cloak (left), what produces a superluminal concentric wave (see 
\cite{UL2011} for a design without supraluminal component).
These modes can propagate in the cloak faster 
than the front wave in vacuum since the frequency dispersion is not introduced in 
the dielectrics, especially those with index values below unity. The cylindrical 
modes excited in the multi-layers then radiate cylindrical waves outside the cloak, 
as evidenced by the right panel in Fig. \ref{fig4}, which explains the  tiny
perturbation of the field observed on right panel of Fig. \ref{fig3} (the
field perturbation is smoothed down at long times, in agreement with the 
analytical part).
In addition, Fig. \ref{fig4} shows a picture of the transient part of the 
field produced by the causal source. Here, we take benefit of the supra-luminal 
propagation of the modes in the cloak to observe that the radiated transient 
part is almost isotropic. We deduce that the radial dependence 
of this transient part does not correspond to 
the function $J_1$ found by A. Sommerfeld and L. Brillouin \cite{Bri1960}, and 
exhibited in the present Eq. (\ref{ETbJ1}). There is no contradiction 
since the $J_1$ dependence is clearly related to 
the Drude-Lorentz model of the dispersion, while the transient field 
around the 2D cloak is related to the effective dispersion produced by the 
cylindrical multilayered geometry. Nonetheless, one can conclude that both 
situations considered in this letter 
attest that the quality of cloaking deteriorates at short times under
illumination by a causal incident field. 

In summary, a new method to analyze propagation of electromagnetic waves in dispersive 
media has been proposed. The major ideas are to consider a layered structure 
to eliminate branch cuts, and an invisible structure (with $\ep = \mu$) to 
eliminate reflections in normal incidence.  
In this situation, the transient regime can be 
highlighted and, especially, an explicit expression
is obtained in the long time limit. As a result the amplitude of 
the transient part decreases like $(t - x/c)^{-3/4}$. Hence the technique
proposed in this letter brings new elements to the method used by Brillouin 
\cite{Bri1960}, where wavefronts (forerunners) can be simply 
exhibited. 
The analysis of the transient regime in the situation of 
the invisible structure has shown that the singularities of the permittivity 
and permeability generate additional contributions to the electric field. 
However, in normal incidence, the contributions vanish 
in the long time limit, thus cloaking is achieved after the 
transient regime. Finally, numerical simulations for a two-dimensional 
cylindrical layered cloak confirm the effect of dispersion, which
affects the quality of cloaking at short times when it is illuminated 
by a causal incident field. 

The proposed method opens new possibilities for investigating transient
regime of dispersive systems, notably structures designed from 
transformation optics like cloaks, carpets, concentrators and rotators. This 
method can be also applied to optical systems moving
at constant relativistic velocity \cite{Thom-Weg2016} and to other wave equations. 
\\

%

\begin{acknowledgments}
B. Avril, A. Diatta and S. Guenneau acknowledge ERC funding (ANAMORPHISM).
G. Arismendi and B. Gralak acknowledge S. Enoch for his support. 
\end{acknowledgments}
%
%



%


\appendix
\section{Supplemental material: calculation of the transient field}
The contribution $E_T^{(b)}(x,t)$ of the two isolated singularities at 
$z = \pm \om_b$ in the integral expression (15) is estimated for 
large values of the relative time $\tau$ [given by (19)] after 
the front wave. These two singularities are present 
in the transmission coefficient $T(z)$ given by (18). 
Decomposing the ratio $z/(z^2-\om_b^2)$ in simple poles, the whole function under the 
integral in (15) can be formulated as
\begin{equation}
f(z) = f_\pm(z) \, 
\exp \left[ - i \, \dfrac{(b-a) \Omega_b^2/(2c)}{z - (\pm \om_b)} \right] \, ,
\label{Tsing}
\end{equation}
where $f_\pm(z)$ are analytic around $\pm \om_b$. Let $\xi = z -(\pm \om_b)$, 
then the functions 
$f_\pm$ and exponential can be expanded in power series around $\xi = 0$:  
\begin{equation}
f(z) = \displaystyle\sum_{q \in \mathbb{N}} 
\dfrac{f_\pm^{(q)}(\pm \om_b)}{q!} \, \xi^q \, 
\displaystyle\sum_{p \in \mathbb{N}} 
\dfrac{[(b-a)\Omega_b^2/(2ic)]^p}{p!} \, \xi^{-p} \, ,
\label{Tseries}
\end{equation}
where $f_\pm^{(q)}(\pm \om_b)$ is the derivative of order $q$ of $f_\pm(z)$ 
evaluated at $\pm\om_b$.
Thanks to the convergence of the series, the terms of this product can be arranged 
in order to obtain the coefficients of the poles $\xi^{-1}$, i.e. the
residues Res$(\pm \om_b)$ of the function $f(z)$ at $z=\pm\om_b$:
\begin{equation}
\text{Res}(\pm \om_b)  = 
\displaystyle\sum_{p \in \mathbb{N}\setminus \{0\}} 
\dfrac{f_\pm^{(p-1)}(\pm \om_b)}{(p-1)!} \, \dfrac{[(b-a)\Omega_b^2/(2ic)]^p}{p!} \, .
\label{Res}
\end{equation}
Notice that it can be checked that the series above converges as well as the 
series expansion of the exponential function. Hence the residues 
$\text{Res}(\pm \om_b)$ are well-defined. 

Using that the complex conjugated of $f(z)$ is $\overline{f(z)} = f(-\overline{z})$, 
the contribution of the singularities at $\pm \om_b$ in the time dependent 
transmitted field is
\begin{equation}
E_T^{(b)}(x,t) = \theta(t - \{ x-x_0 + \alpha -a \} / c ) \, \text{Imag} \, 
\Big\{ 4 \pi \text{Res}(\om_b) \Big\} \, .
\label{Eres}
\end{equation}
The exact calculation of this second contribution, corresponding to the 
transient regime, cannot be performed in general. However, the  
$(x,t)$ dependence can be analyzed from the one of $f_\pm(z)$ 
which can be expressed as 
\begin{equation}
f_\pm(z) = g_\pm(z) \exp[-i z \tau] \, , \quad 
\tau = t - (x-x_0 + \alpha -a ) / c \, .
\end{equation}
where the functions $g_\pm(z)$ are $(x,t)$ independent, and the time quantity 
$\tau$ defines the arrival of the signal (from $\tau = 0$). Denoting 
$\beta = (b-a) \Omega_b^2/(2\om_b^2 c)$ and recalling that $\xi = z -(\pm \om_b)$, 
the function (\ref{Tsing}) becomes
\begin{equation}
f(\xi \pm \om_b) = g_\pm(\xi \pm \om_b) \exp[\mp i \om_b \tau] \, 
\exp[-i( \tau \xi + \om_b^2 \beta / \xi) ] . 
\end{equation}
Then the residues can be expressed as 
\begin{equation}
\text{Res}(\pm \om_b) = \dfrac{1}{2 i \pi} 
\displaystyle\int_{|\xi| = d} d \xi \, f(\xi \pm \om_b)
\label{residus}
\end{equation}
as soon as the functions $g_\pm(z)$ are analytic in the disks of radius $d$ 
and centered at $\pm \om_b$. In particular, this expression can be estimated 
for $\tau$ tending to infinity. Let the radius of the disks set to 
$d = \om_b \sqrt{\beta/\tau}$ and the complex number 
$\xi = \om_b \sqrt{\beta/\tau} \exp[i \phi]$. For $\tau / \beta \rightarrow \infty$, 
the functions $g_\pm(\xi \pm \om_b) \approx g_\pm(\pm \om_b)$ and 
the residues can be approached by 
\begin{equation}
\begin{array}{ll}
\text{Res}(\pm \om_b) & \approx
\dfrac{1}{2 i \pi} g_\pm(\pm \om_b) \exp[-i (\pm \om_b) \tau] \, 
i \om_b \sqrt{\beta/\tau} \\[2mm]
& \times \displaystyle\int_{[0,2\pi]} d \phi \exp[i \phi - i 
2 \om_b \sqrt{\beta \tau} \cos \phi] \, . 
\end{array}
\label{Res-tau}
\end{equation}
Using the integral representation of the Bessel function 
\begin{equation}
J_1(u) = - \, \dfrac{1}{2 i \pi} \displaystyle\int_{[0,2\pi]} 
d \phi \exp[i \phi - i u \cos \phi] \, , 
\label{J1}
\end{equation}
it is deduced that, for $\tau / \beta \rightarrow \infty$, 
\begin{equation}
\begin{array}{ll}
\text{Res}(\pm \om_b) \approx & - \, i \, \dfrac{S_0 \mu_0 c}{2} \, 
\dfrac{\om_0 \om_b }{\om_b^2 - \om_0^2} \, \sqrt{\beta/\tau} \\[2mm]
& \times \exp[\mp i \om_b (\tau+\beta/2)] 
\, J_1(2 \om_b\sqrt{\beta \tau}) \, . 
\end{array}
\label{Res-J1}
\end{equation}
Replacing this estimate of the residues in (\ref{Eres}) provides the expression 
(20) of the time dependent transmitted field in the letter. 
\end{document}